# SPAD-based asynchronous-readout array detectors for image-scanning microscopy

**Mauro Buttafava[1], Federica Villa[1], Marco Castello[2], Giorgio Tortarolo[2,3], Enrico Conca[1], Mirko Sanzaro[4], Simonluca Piazza[5], Paolo Bianchini[5], Alberto Diaspro[5,6], Franco Zappa[1], Giuseppe Vicidomini[2] and Alberto Tosi[1,\*]**

[1]*Dipartimento di Elettronica, Informazione e Bioingegneria, Politecnico di Milano, Milano (Italy).*
[2]*Molecular Microscopy and Spectroscopy, Istituto Italiano di Tecnologia, Genova (Italy).*
[3]*Dipartimento di Informatica, Bioingegneria, Robotica e Ingegneria dei Sistemi, Università di Genova, Genova (Italy).*
[4]*Was with 1 during this research activity.*
[5]*Nanoscopy and NIC@IIT, Istituto Italiano di Tecnologia, Genova (Italy).*
[6]*Dipartimento di Fisica, Università di Genova, Genova (Italy).*
*\*alberto.tosi@polimi.it*

**Abstract:** Fluorescence microscopy and derived techniques are continuously looking for photodetectors able to guarantee increased sensitivity, high spatial and temporal resolution and ease of integration into modern microscopy architectures. Recent advances in single-photon avalanche diodes (SPADs) fabricated with industry-standard microelectronic processes allow the development of new detection systems tailored to address the requirements of advanced imaging techniques (such as image-scanning microscopy). To this aim, we present the complete design and characterization of two bidimensional SPAD arrays composed of 25 fully independent and asynchronously-operated pixels, both having fill-factor of about 50% and specifically designed for being integrated into existing laser scanning microscopes. We used two different microelectronics technologies to fabricate our detectors: the first technology exhibiting very low noise (roughly 200 dark counts per second at room temperature), and the second one showing enhanced detection efficiency (more than 60% at a wavelength of 500 nm). Starting from the silicon-level device structures and moving towards the in-pixel and readout electronics description, we present performance assessments and comparisons between the two detectors. Images of a biological sample acquired after their integration into our custom image-scanning microscope finally demonstrate their exquisite on-field performance in terms of spatial resolution and contrast enhancement. We envisage that this work can trigger the development of a new class of SPAD-based detector arrays able to substitute the typical single-element sensor used in fluorescence laser scanning microscopy.

## 1. Introduction

Since the early 90s, single-photon detectors started to play a growing role in scientific and industrial environments, and microscopy was one of the first applications taking advantage of the extremely high sensitivity of these devices. Confocal laser-scanning microscopy (CLSM) [1], fluorescence lifetime image microscopy (FLIM) [2], fluorescence correlation spectroscopy (FCS) [3], down to modern super-resolution techniques, such as stimulated emission-depletion (STED) [4, 5] and image-scanning microscopy (ISM) [6, 7], are only some of the applications currently enabled by single-photon detectors.

Vacuum-based devices, like photomultiplier tubes (PMT) and micro-channel plates (MCP), historically played a leading role in the field, despite limitations like fragility, intrinsic deterioration with usage, high cost, bulkiness and operation complexity. This was essentially due to advantages like large active area and high measurement dynamic range.

More recently, microelectronic single-photon detectors like SPADs (single-photon avalanche diodes) are gaining importance in microscopy applications thanks to their great reliability, large robustness, ease of operation, high detection efficiency, low timing jitter and their integrability with read-out circuits, allowing for the development of arrays. SPADs are essentially p-n junctions reversed-biased above their breakdown voltage [8, 9], where absorbed photons can generate a self-sustaining carrier multiplication process (avalanche), which eventually translates into a macroscopic current that can be easily detected by an external discriminator circuit. Silicon SPADs are used to detect photons in the visible wavelength range, from 400 nm to 1000 nm, while SPADs based on III-V compound materials (like InGaAs/InP) are useful to detect signals in the near-infrared region (from 900 nm to 1700 nm) [10].

Single-pixel silicon SPADs can be divided into two categories, according to their internal device structure: i) the so-called *thick SPADs* [11, 12] are commonly employed in fluorescence microscopes thanks to their very good photon detection efficiency (PDE), which can be as high as 70% at 780 nm, but are characterized by a poorer temporal resolution (350 ps, FWHM) with respect to the so-called *thin SPADs* [13, 14, 15], which can reach temporal resolutions down to 30 ps at the expense of a lower PDE. Besides the higher temporal resolution (of fundamental importance in time-resolved microscopy applications like FLIM), the real strength of thin SPADs is the compatibility with microelectronic circuits, which allows the integration of dedicated electronics into the same silicon chip [16] for creating 1-D and 2-D arrays of detectors, effectively implementing single-photon imagers with photon timing capability.

A wide variety of monolithic SPAD array implementations can be found in the literature [17]. Suitable fabrication technologies include CMOS process nodes, down to 40 nm [18], as well as lower-density, but more consolidated nodes, e.g. 0.35 µm [19]. New opportunities are also offered by the recently explored 3D-stacked imagers, employing two different technologies for the detector array and the front-end electronics [20, 21]. Pixel number can be as high as 512-by-128 [22] and in-pixel electronics can include up/down counters [23], time-to-digital converters [18, 24, 25], time-gating [25] and coincidence detection circuits [25, 26]. However, the combination of high pixel-number and embedded processing circuits has the drawback of generating massive quantity of data, that needs to be transferred outside the chip. Image readout is then usually implemented using serial communication protocols and is based on frames (i.e. data related to the entire imager is downloaded periodically, independently of the number of triggered pixels). The frame-rate is usually limited to few hundreds of kilo-frames/s, also depending onto the communication interface used to transfer data from the detection system to the PC (typically USB 2.0 or 3.0). This approach can be a bottleneck in applications not requiring high pixel-number, but rather fast readout speed and the possibility to independently address each pixel, such as ISM.

In a nutshell, image scanning microscopy requires to collect the image of excitation/detection region for each scanning position of the sample. Since the excitation region is diffraction-limited, its size is typically in the range of few hundreds nanometer (the size reduces when ISM is combined with STED microscopy [27]) and the pixel dwell-time is in the microseconds range (few tens of nanoseconds when ISM is combined with resonant scanners). Thus, it is clear the importance of having a detector array with a limited number of pixel but with asynchronous readout (no frame-rate). The lack of a detector which this requisite was the major reason for the delay in the practical implementation of ISM. Indeed, ISM has been proposed in the 80s [28, 29], offering an approach to solve one of the most critical aspect of CLSM (effectively introducing it in the pool of super-resolution techniques) but it took more than 30 year to see a versatile implementation of this idea. In fact, confocal laser scanning microscopy can be considered a super-resolution microscopy technique, allowing the diffraction barrier to be overcome by a factor of $\sqrt{2}$, as defined by the full-width at half-maximum (FWHM) of the point spread function. In practice, however, this improvement can be obtained only by reducing the diameter of the confocal pinhole, but this translates into a significant reduction of the signal-to-noise ratio (SNR) of the resulting images.

Since in ISM the single-pixel photodetector is replaced by an imaging detector and the pinhole is removed (or opened to a size major of 1 Airy unit), each pixel/element of the detector effectively acts a virtual pinhole, but all the light reaching the image plane is collected. The final ISM image is then obtained by computationally combining the information contained in the 2D dataset of the acquired images [30, 31], i.e. after each scanning the microscope produces one confocal image for every pixel/element of the detector array (the so-called scanned images). The most important aspect of ISM is the increase of the detected signal level compared to simple confocal imaging (thus allowing imaging with lower excitation power), while at the same time slightly improving upon the confocal resolution limit.

ISM implementations have been demonstrated with conventional cameras [6], at the expenses of a low imaging speed (due to the limited camera frame-rate) or with *optomechanical* implementations [32, 33, 34, 35], which have the drawback of a substantial modification of the microscope structure. Recently, the imaging speed limit was addressed by the *AiryScan* implementation of CLSM [36], for which a 2D bundle of optical fibers is coupled to a linear array of GaAsP PMTs. However, this solution still hinders the temporal information related to photon arrivals (thus preventing the implementation of FLIM and similar techniques) and exhibits some of the typical restrictions of vacuum-based detectors (like fragility and high cost).

It is thus clear that designing a small SPAD array with picosecond timing-ability, fully independent pixel operation and data readout would be a very effective solution to overcome all the above limitations, and to implement faster and more flexible image-scanning microscopes.

In this work, we describe design and characterization of two 5-by-5 SPAD arrays, fabricated in two different technologies and specifically tailored for ISM applications. Theoretical studies show that this relatively small number of pixels (i.e., 25) is sufficient for practical applications, since a higher number of elements would provide only a marginal spatial resolution gain [31]. For both implementations, in the following sections we will describe the SPAD structures, the overall imager architectures, their optical and electrical characterization and an example of ISM used for imaging the convoluted tubulin network of a human cell.

## 2. Sensor design

The microelectronic fabrication processes used for our asynchronous-readout image sensors are: i) a 0.35 µm High-Voltage CMOS technology (0.35 µm–HVCMOS) [19] and ii) a 0.16 µm Bipolar-CMOS-DMOS technology (0.16 µm–BCD) [37]. Sensor architecture and geometry remain the same, but the detection performance is strongly affected by the chosen technology. In detail, devices developed using our 0.35 µm–HCMOS are characterized by best-in-class dark-count noise, while 0.16 μm–BCD SPADs feature bespoke dopant implants for enhanced detection efficiency.

### 2.1 SPAD fabrication technologies

A simplified cross-section of our SPADs fabricated using the 0.35 µm–HVCMOS technology is shown in Fig. 1(A). A high-voltage n-well isolates the device from the p-type substrate. Shallow p+ and n+ implants are the anode and cathode contacts of the device, respectively. A further low-energy n-type enrichment implant defines the high field multiplication region (i.e. the SPAD active area). Finally, a p-type implantation is used as a guard-ring to suppress premature edge breakdown effects. Although a common cathode n-well could be shared by all SPADs of the array enhancing the fill-factor, the SPAD pixels reported in this paper are isolated from one another to avoid electrical crosstalk. In this device structure the avalanche is mainly triggered by holes, which have an impact ionization coefficient lower than electrons, thus leading to a lower PDE.

The simplified cross-section of the 0.16 µm–BCD SPAD [37] is shown in Fig. 1(B). Each device is fully enclosed in a double-well pocket, formed by a n-type buried layer, for isolation from the p-type substrate, and a heavily doped n-type well, which provides a low resistance path to the cathode contact. This fabrication technology also features deep trenches, which here are exploited for electrical and optical isolation between pixels. As in the 0.35 µm–HVCMOS SPAD, a bespoke enrichment implant defines the avalanche region. However, in this case, a high-energy p-type implant is used. As a result, the avalanche region has been moved towards the n-type buried layer and the avalanche current is mainly triggered by electrons.

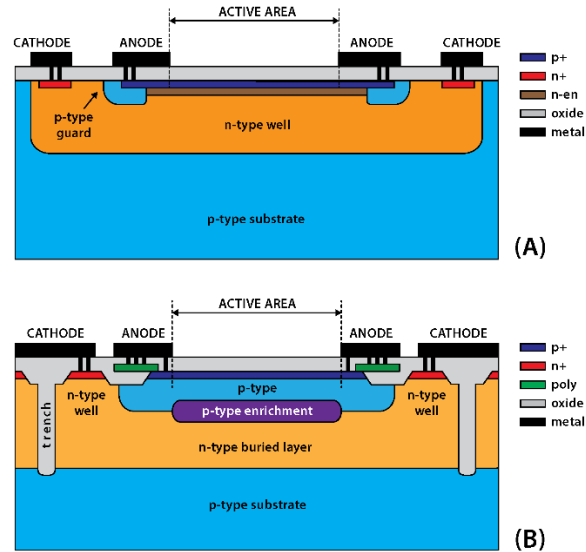

Fig. 1 Simplified cross-section of the SPAD inside each individual imaging pixel, fabricated using the 0.35 µm–HVCMOS (A) and the 0.16 µm–BCD SPAD technology (B). Both fabrication processes are industry-standards and allow for the integration of in-pixel electronics (not shown in the cross-section). Features in the images are not in scale.

*2.2 Array geometry*

To allow for fully-asynchronous and independent readout from the image sensor, while maintaining the maximum flexibility in data processing, the output signal from each pixel is routed to a digital output pad. Each photon detection in the $N^{th}$ pixel is marked by the trailing edge of a voltage pulse onto the $N^{th}$ output pad, with a time uncertainty (jitter) lower than few tens of picoseconds. The drawback is the growing complexity when increasing the pixel number, due to longer electrical connections and the higher pin-count. The square geometry of 5-by-5 pixels was chosen as a compromise between spatial resolution and device complexity, always keeping in mind the target ISM application, while different implementations available in literature make use of hexagonal patterns [38]. Table 1 summarizes the geometry details of the designed sensors.

Table 1: Image sensors single-pixel geometry details and overall fill-factor.

|  | 0.35 µm–HVCMOS | 0.16 µm–BCD |
|---|---|---|
| Pixel side length | 50 µm | 57 µm |
| Pixel corner radius | 5 µm | 5 µm |
| Pixel pitch | 75 µm | 75 µm |
| Array fill-factor | 44 % | 57.5 % |

The pixel active area is square with rounded corners, in order to maximize the array fill-factor (i.e. the ratio between photosensitive area and overall silicon area). A 5 µm curvature radius is sufficient to avoid premature edge-breakdown effects due to electrical field peaking at corners. The pixel pitch (i.e. distance from center to center) is 75 µm, with a side length of 50 µm for the 0.35 µm-HVCMOS device (leading to a fill-factor of about 44 %), which is increased to 57 µm for the 0.16 µm–BCD device, thanks to the smaller minimum feature-size of this technology (which allows for a smaller gap between pixels, leading to a fill-factor of around 57.5 %). A detail of the photosensitive section of both devices is shown in Fig. 2(A, B). The overall image sensors have total dimensions of $2.2 \times 2.4$ mm$^2$ (mainly limited by the output pads) for the 0.35 µm-HVCMOS one and $1.3 \times 1.2$ mm$^2$ (owing to the smaller pad pitch) for the 0.16 µm–BCD one. In this context, it is important to highlight that having a relatively small total active area (i.e. ~ $350 \times 350$ µm$^2$) is convenient from an optical point of view, since having a projected size of the SPAD array on the sample plane of ~ 1 Airy unit requires to add only a relatively small (i.e. 1-10×) extra magnification to a conventional laser scanning architecture.

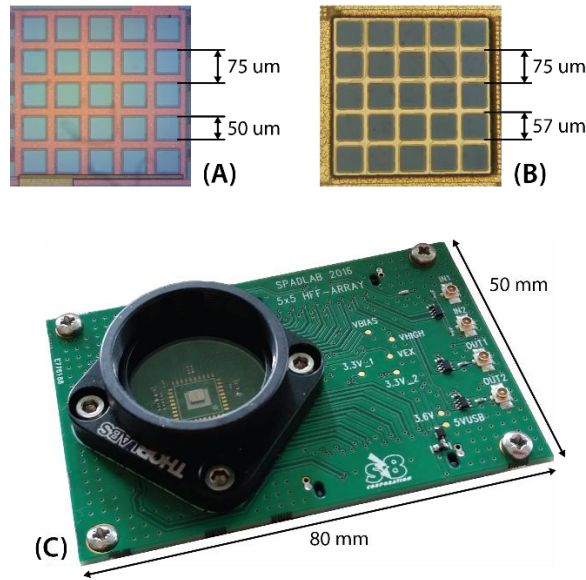

Fig. 2 Detail of the photosensitive section of the 0.35 µm-HVCMOS (A) and the 0.16 µm–BCD (B) imagers respectively, showing the 5-by-5 square SPAD array. (C) The frontend board (is part of the complete detection system), hosting the image sensor and dedicated electronics.

*2.3 Readout and quenching circuit*

A SPAD is able to detect photons when its reverse bias voltage $V_{BIAS}$ is raised above the breakdown value (which is around $V_{BD}$ = 25 V for both technologies). The difference between bias and breakdown voltages is called excess-bias voltage ($V_{EX} = V_{BIAS} - V_{BD}$) and its value has a strong impact on SPAD performance, as shown later. After each photon detection, the avalanche current has to be sensed by an external circuit and the bias voltage must be reduced below breakdown as quickly as possible, in order to *quench* the avalanche [9]. Correspondingly, a low-jitter output pulse is generated to mark the photon arrival time. After each avalanche quenching, the SPAD is kept disabled (i.e. biased below $V_{BD}$) for few tens of nanoseconds (the so-called *hold-OFF* phase) in order to lower the probability of afterpulses [39]. Finally, the device is rearmed raising its bias voltage back to $V_{BD} + V_{EX}$.

Both imagers employ similar architectures for detector activation and avalanche readout. Each SPAD is connected to an independent active-quenching circuit based on a Variable-Load Quenching Circuit implementation (VLQC) [40], shown in the diagram of Fig. 3. When the SPAD is ready to detect photons, the transistor $M_Q$ is weakly turned ON (showing a series resistance of few kilo-ohms) and the detector voltage is raised above breakdown. Transistors $M_R$ and $M_G$ are both turned OFF. When a photon is detected, the avalanche current flows through $M_Q$ and the resulting voltage drop is sensed by the control logic through the SENSE input. Then the control logic completely turns $M_Q$ OFF, thereby increasing its impedance, and quickly turns $M_G$ ON, thus quenching the avalanche current by pulling up the anode voltage to $V_{EX}$. Correspondingly, a digital voltage pulse is generated at the EVENT OUT pin. After, the hold-OFF phase is enforced, keeping that pixel disabled for the entire hold-OFF time ($T_{HO}$). To this aim, a common externally provided analog voltage ($V_{HO}$) is used to set the duration of $T_{HO}$ for all the pixels. At the end of this phase, $M_G$ is turned OFF, $M_Q$ is weakly turned ON again and the transistor $M_R$ is briefly activated, in order to force back the anode voltage to ground, thus restoring the original SPAD bias conditions in less than 1 ns.

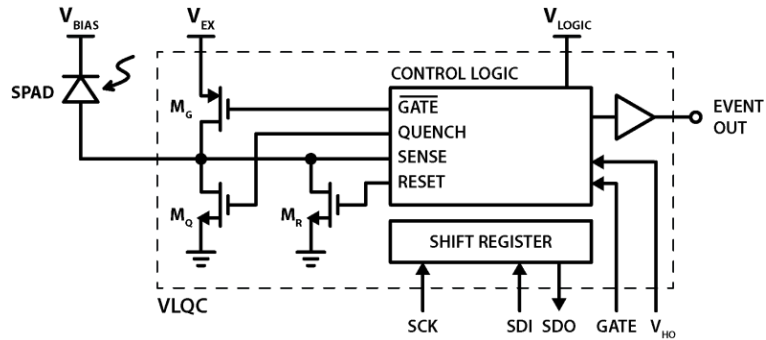

Fig. 3 Simplified circuit diagram of the in-pixel readout and quenching logic, based on a Variable-Load Quenching Circuit. Each pixel operates independently, marking photon detections by a voltage pulse at the output and subsequently enforcing the hold-off phase with programmable $T_{HO}$ duration.

Additionally, a global enable input (GATE) allows to simultaneously de-activate all SPADs by connecting their anodes to the $V_{EX}$ supply rail, thus biasing them below the breakdown voltage. Each individual pixel can also be turned ON/OFF via a configuration serial interface, allowing for exclusion of those SPADs whose outputs are not of interest for the measurement (thus preventing the increase of noise in neighbor pixels due to optical and electrical crosstalk) or which are too noisy. This serial interface is composed by a common clock signal (SCK) and a daisy-chained data line (SDI for input, SDO for output) passing through the 25 pixels.

The frontend makes use of thick-oxide transistors ($M_Q$, $M_R$, $M_G$) capable of withstanding the high excess bias of the SPADs, while the rest of the sensing circuitry is made with low voltage transistors (3.3 V and 1.8 V for the HVCMOS and BCD chips, respectively). The readout and quenching circuitry are placed just outside the imaging area in order to maximize its fill-factor.

### 2.4 Detection System Design

A complete and standalone detection system was developed in order to easily characterize the two image sensors and to exploit them in our ISM setup. It is based on two stacked Printed Circuit Boards (PCBs). The upper one, called *frontend board*, is shown in Fig. 2(C) and hosts: (i) the detector; (ii) the bias voltage generator (implemented using a switching-mode boost converter followed by a linear voltage regulator); (iii) the hold-OFF time control DAC (Digital-to-Analog Converter) and (iv) the serial communication interface used to set the enabled/disabled status of each array pixel. The sensor chip is directly mounted onto the PCB and is electrically connected through direct bonding wires (Chip-On-Board mounting technique). A SM1-threaded mechanical mounting flange (*Thorlabs Inc.*) can be used to optically couple the system to the experimental setup.

Through a pair of high-density connectors, the *frontend board* is connected to a second PCB, called *connection board*, which contains: (i) the global power supply section, used to generate the system voltage rails (5.0 V, 3.3 V and 1.8 V) starting from the common external 5 V – 1 A supply; (ii) an 8-bit microcontroller, used to control and manage the entire system, and (iii) a set of 25 low-jitter buffers, able to drive 50 Ω impedance cables. Each buffered output, providing 3.3 V voltage pulses synchronous to photon detections, is connected to a coaxial cable with SMB (Sub-Miniature, type B) connectors. Cable connections for the serial interface control lines are also provided, allowing the user to independently enable each pixel.

### 3. Experimental characterization

Before integration into the microscope, both SPAD arrays have been fully characterized in terms of detection efficiency, noise, afterpulsing probability, optical crosstalk and timing jitter.

### 3.1 Photon detection efficiency

Photon Detection Efficiency (PDE) of the two imagers has been measured at a temperature of 300 K, over a wavelength range between 400 nm and 1000 nm, obtaining the results shown in Fig. 4(A). The detector fabricated using the 0.35 µm-HVCMOS technology (blue line) has a peak PDE of 40% at 430 nm which drops below 10% starting from 750 nm, as already seen in similar devices reported in [19]. The 0.16 µm–BCD sensor (red line) shows a substantially higher PDE, having a peak value of 65% around 500 nm and remaining above 20% in the entire range from 400 nm to around 750 nm. The excess bias voltage was set to 6 V and 5 V for the 0.35 µm-HVCMOS and 0.16 µm–BCD devices respectively, as the optimal trade-off values between detection performance metrics. As expected, owing to the higher avalanche triggering probability of electrons [37], the 0.16 µm–BCD SPAD leads to higher PDE compared to the 0.35 µm–HVCMOS ones (where avalanches are triggered by holes) at all wavelengths. The interference ringing shown by both curves is due to the multiple dielectric layers and interfaces deposited on the chips during the Back-End-of-Line (BEoL) production phase. As a reference, Fig. 4(A) also includes the PDE of a GaAsP photomultiplier tube [41] (green line), which is commonly used as detection element in fluorescence laser scanning microscopy. The 0.16 µm–BCD imager shows superior detection performance across the full measured wavelength range.

The uniformity of PDE inside the entire active area of a single pixel (the central one) was also measured, thanks to a laser-point scanning system, for both fabrication technologies. The resulting (normalized) 2-D count maps are reported in Fig. 4(B, C), showing extremely good uniformity and sharp drops outside active area borders, demonstrating a good electric field uniformity and the absence of edge peaking.

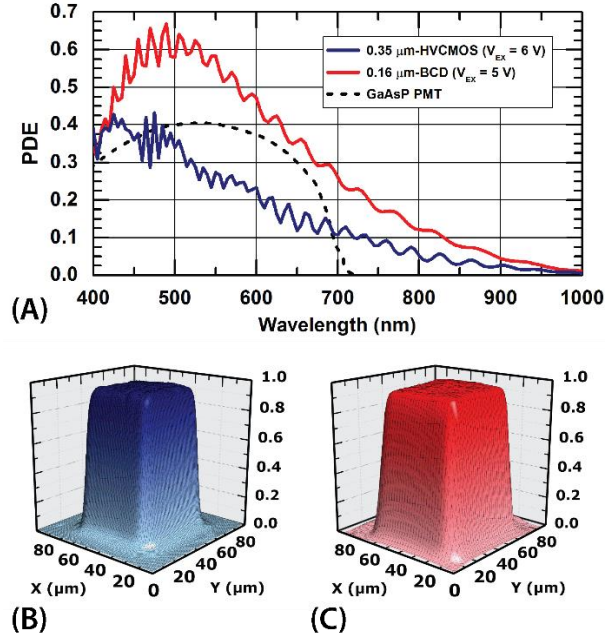

Fig. 4 (A) Photon detection efficiency of 0.35 µm–HVCMOS (blue curve) and 0.16 µm–BCD (red curve) sensors, measured in the 400 nm – 1000 nm wavelength range. As a comparison, PDE of a GaAsP PMT is also reported [41] (green curve). (B, C) Normalized PDE uniformity inside the active area (for an individual pixel) of the 0.35 µm–HVCMOS and the 0.16 µm–BCD sensors, respectively.

### 3.2 Noise

SPAD noise can be identified as all the electrical output pulses that are not due to photon detections. There are two major phenomena contributing to SPAD noise. The first one is related to avalanche ignitions triggered by carriers due to either thermal generation processes [42] or trap-assisted tunneling [43]. Such average rate is the SPAD dark count rate (DCR). The second noise contribution is due to avalanches triggered by carriers that may get trapped by deep levels during a previous avalanche and are released with a stochastic delay, eventually igniting a so-called *afterpulse* when the SPAD is re-armed [39]. This effect is quantified by the SPAD afterpulsing probability. Besides material quality, DCR depends also on device design and fabrication, operating temperature and applied excess-bias voltage. On the other hand, afterpulses are strongly correlated to the detected signal, thus causing a non-linear distortion of the acquired data. The afterpulsing effect is mitigated keeping the device OFF for a long time (called hold-OFF time, $T_{HO}$), typically tens of nanoseconds (from 20 ns to 200 ns), after each photon detection and indeed limits the maximum counting-rate of each array element/pixel (with an asymptotical limit equal to $1/T_{HO}$).

Fig. 5 shows the percentage distribution of DCR (related to individual pixels) for the two imagers herein described. The 0.35 µm–HVCMOS SPAD (blue line) exhibits a DCR median of 200 counts per second (cps) at ambient temperature (300 K) with 6 V of excess-bias voltage. The yield of this production technology can be inferred looking at the knee of the curve, where the measured DCR starts to rise above the median value. It can be estimated around 75-80%. However, the threshold above which a pixel is considered a so-called *hot-pixel* (i.e. having a too-high DCR to be usable) strongly depends on the application and for ISM it can be easily considered above few kcps. The 0.16 µm–BCD SPAD have a higher DCR, with a median value of 2 kcps at 300 K and 5 V of excess-bias. Here it is harder to define the yield since there is a constant increase of DCR, at least from 30% up. In this case, due to the higher DCR, a proper testing and a careful selection of the device is required for each array to be used in ISM measurements.

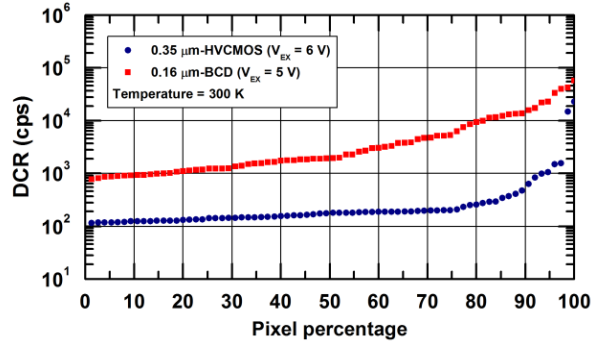

Fig. 5 Percentage distribution of dark count rates (related to individual pixels) for the two imagers. The 0.35 µm-HVCMOS SPADs (blue curve) have a DCR median of 200 cps at 300 K and 6 V of excess-bias, while 0.16 µm–BCD ones (red curve) have a higher DCR median of 2 kcps at 300 K with 5 V excess-bias.

The afterpulsing probability at various hold-OFF times ($T_{HO}$) is reported in Table 2. It has been measured by recording into a histogram the inter-arrival times between consecutive output pulses of an individual pixel. The contribution of simple DCR to this histogram can be fitted with an exponential decay at long inter-arrival times and then subtracted from the experimental data in order to have only the contribution of avalanches due to afterpulses. Their probability is then computed as the integral sum of afterpulsing events, divided by the integral sum of the histogram itself (i.e. the total number of avalanches).

The afterpulsing performance of the 0.16 µm–BCD is outstanding, showing only a 0.31% probability with $T_{HO}$ as short as 25 ns (equivalent to a maximum count-rate of 40 Mcps per pixel). On the other side, the minimum hold-OFF time needed to operate 0.35 µm–HVCMOS SPADs with negligible afterpulsing effects (< 3%) rises to 100 ns (equivalent to 10 Mcps of maximum count-rate per pixel).

Table 2: Afterpulsing probability at various hold-OFF times
for 0.35 µm–HVCMOS and 0.16 µm–BCD SPAD arrays.

| Hold-OFF time ($T_{HO}$) | 0.35 µm–HVCMOS | 0.16 µm–BCD |
| --- | --- | --- |
| 25 ns | 14.70% | 0.31% |
| 50 ns | 5.33% | 0.25% |
| 100 ns | 2.37% | 0.18% |
| 200 ns | 1.59% | 0.09% |

*3.3 Temporal response*

The temporal response of a SPAD is very important for time-resolved applications, like fluorescence-lifetime ISM (FLISM), i.e., the combination of FLIM and ISM. Notably, FLISM, to the best of our knowledge, is the only effective super-resolution FLIM technique. The temporal response of a SPAD can be measured illuminating the device by means of a narrow pulsed laser (having a width of few tens of picoseconds) and acquiring the distribution of photon arrival times by means of the time-correlated single-photon counting (TCSPC) technique [2]. A typical SPAD temporal response is composed by a narrow peak and a subsequent slower exponential tail [44]. The peak is essentially due to photons directly absorbed in the depleted region, where a photogenerated electron-hole pair is immediately separated, thus trigging the avalanche. Its FWHM gives a good indication of the detector timing resolution. The slower exponential tail is due to photons that have been absorbed within the device neutral regions. These carriers can diffuse and finally reach the high field region with some probability (many of them recombine before reaching it), eventually triggering a delayed avalanche with respect to the photon absorption time.

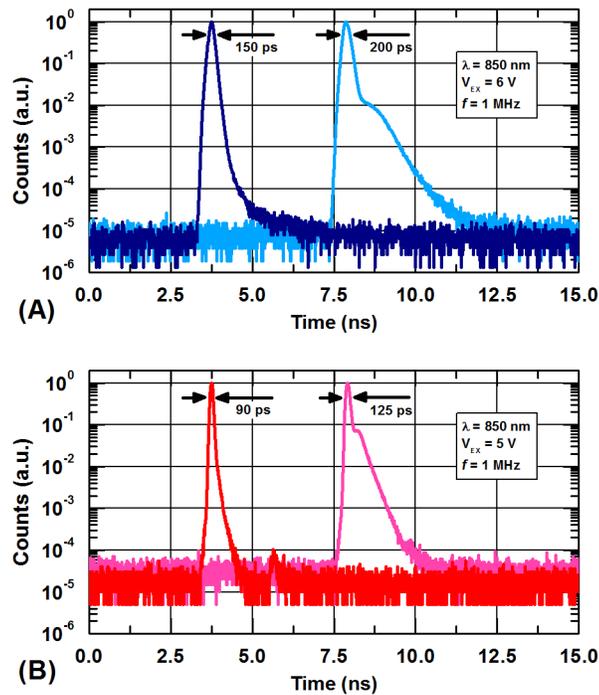

Fig. 6 Single-pixel temporal response of 0.35 µm–HVCMOS (A) and 0.16 µm–BCD (B) imagers, measured using a 50 ps FWHM pulsed laser at 850 nm. The temporal responses have been measured both with only one pixel turned ON (blue and red curves) and with the all 25 pixels simultaneously ON and illuminated (cyan and magenta curves), where is clearly visible the effect of optical crosstalk between adjacent pixels.

Fig. 6(A) shows the single-pixel temporal response of the 0.35 µm–HVCMOS imager: the blue curve is acquired with the remaining 24 pixels turned OFF, while the cyan curve is acquired with all the 25 pixels turned ON and illuminated. The characterization is performed using a pulsed diode laser at 850 nm, having less than 50 ps FWHM width and working at 1 MHz repetition rate (*Advanced Laser Diode System GmbH*). Photon arrival times were acquired using an SPC-630 TCSPC board (*Becker&Hickl GmbH*) with a time jitter of less than 8 ps

FWHM. With all the remaining pixels turned OFF, the single-pixel temporal response width is 150 ps FWHM, with a fast exponential tail of 60 ps time-constant. The measured temporal response is consistent across all 25 elements. When turning ON and illuminating the entire array, the single pixel temporal response width rises to 200 ps FWHM, due to electrical crosstalk between switching signals (either inside the chip itself and through the frontend board). Furthermore, an additional bump appears after the main peak, due to photons generated by optical crosstalk between adjacent pixels (see section 3.4).

Fig. 6(B) shows the temporal response of the 0.16 µm–BCD sensor. With only one pixel turned ON (red curve), the response width is narrower than 90 ps FWHM, with an exponential tail time-constant of 50 ps. With all the 25 pixels turned ON (magenta curve), the response width slightly increases to 125 ps FWHM and, also in this case, the effect of optical crosstalk becomes visible as an additional bump.

*3.4 Optical crosstalk*

Carriers flowing inside a SPAD during each avalanche can cause the emission of secondary photons, due to hot-carrier relaxation phenomena. These secondary photons, propagating throughout the chip, can be absorbed into the active region of a nearby device, eventually causing spurious avalanches and degrading the measurement SNR. This effect is known as optical crosstalk [45] and is influenced by several factors like: (i) device material and structure; (ii) distance between neighbor pixels (i.e. array pitch); (iii) intensity and duration of the avalanche current; (iv) PDE of each pixel. The optical crosstalk probability can be quantified by measuring the temporal correlation between photon arrival times of two neighbor pixels, under weak ambient light. In absence of optical crosstalk phenomena, the inter-arrival times distribution (i.e. the distribution of time differences between a first event on one channel and a second event on the other channel, and vice-versa) should follow an exponential decay (being a combination of two uncorrelated Poissonian processes). Crosstalk events create a variance from this theoretical trend and can be numerically quantified subtracting the latter from acquired data and normalizing.

Table 3 summarizes crosstalk probability values for the presented detectors, relative to first neighbor pixels, both in the orthogonal and diagonal directions. As anticipated, despite the relatively small probabilities (1.5% for the 0.35 µm–HVCMOS and 5% for the 0.16 µm–BCD sensors), the effect of optical crosstalk is clearly visible also looking at the single-pixel temporal response curves when all the pixels are turned ON and illuminated (Fig. 6, cyan and magenta curves). Notwithstanding the presence of deep oxide trenches, crosstalk probability is higher for the 0.16 µm–BCD for multiple reasons: (i) smaller pixel separation, (ii) higher PDE, (iii) higher intensity of the avalanche current and (iv) multiplication region is deeper than the oxide trenches [46]. However the relatively higher crosstalk probability of the 0.16 µm–BCD sensor does not introduce any degradation in the spatial gain resolution (with respect to CLSM) obtained via ISM.

**Table 3: Optical crosstalk probability between adjacent pixels (orthogonally and diagonally) for both imagers.**

| Pixel position | 0.35 µm–HVCMOS | 0.16 µm–BCD |
|---|---|---|
| First neighbors (orthogonal) | < 1.5 % | < 5 % |
| First neighbors (diagonal) | < 0.2 % | < 0.3 % |

## 4. Image scanning microscopy experiments

Our imagers have been successfully integrated into ISM experiments, to demonstrate the advantages of a SPAD-based detector and to compare the lower noise of 0.35 µm-HVCMOS device against the higher PDE of 0.16 µm-BCD one. Both SPAD arrays have been integrated into a custom confocal laser-scanning microscope, replacing its single-point detector, as described in detail in [47, 48, 27]. Output lines from the sensor array have been connected to an FPGA-based board (*NI-USB-7856R from National Instruments*) for counting photons detected in each laser spot position and for managing the entire microscope system (including the synchronization with scanning devices). Measurement control, data-acquisition and image reconstruction are performed using the *Carma* custom software [49, 47].

Fig. 7 shows images of tubulin filaments in a Hela cell stained with Abberior STAR red, where the 25 raw scanned images acquired from each independent pixel are processed using the adaptive pixel-reassignment (APR) method of [31], for obtaining the final ISM images. Pixel dwell-time is 100 µs and the imaged sample area is equal to 10 x 10 µm$^2$. The photobleaching effect on sample is reduced by using a low laser excitation power ($P_{EX}$), at the expense of a small amount of fluorescence photons. Exacerbating this condition, as shown in Fig. 7(A, B) where $P_{EX}$ = 17 nW, the contrast is higher in the image obtained with the 0.35 µm-HVCMOS sensor, owing to its lower DCR. However, the lower PDE of 0.35 µm-HVCMOS SPADs may translates into the loss of some details of the sample structure in figure Fig. 7(A) respect than Fig. 7(B). Increasing the excitation power to $P_{EX}$ = 110 nW translates into better image quality for both SPAD technologies, Fig. 7(C, D). However, the higher PDE and fill-factor of the 0.16 µm-BCD imager lead to a 4x signal improvement (543 photons against 152 photons), thus an higher SNR.

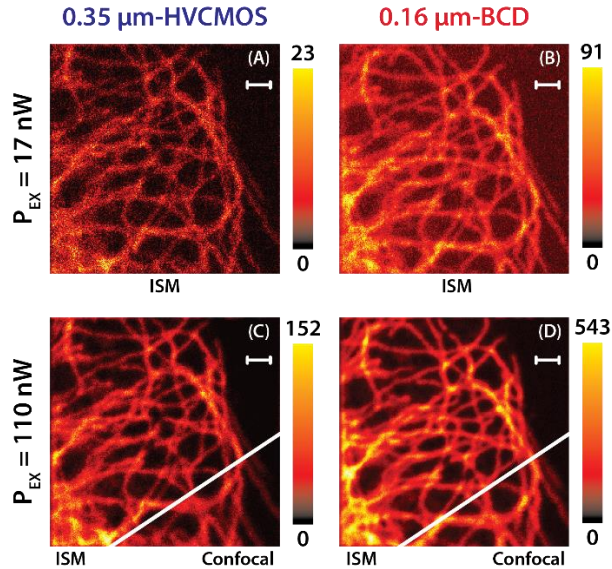

Fig. 7 ISM images of tubulin filaments stained with Abberior STAR red, acquired using the two described 5-by-5 SPAD imagers and processed with the Adaptive Pixel Reassignment (APR) method discussed in [31]. Pixel dwell-time: 100 µs. Sample area: 10 x 10 µm$^2$ (scale bar: 1 µm). As a comparison, (C) and (D) also show the difference between ISM and standard confocal images (obtained simply summing together data from all the 25 pixels).

## 5. Discussion

The 0.35 µm-HVCMOS was the earlier among the two imagers to be designed and fabricated and allowed us to successfully demonstrate for the first time the advantages of using asynchronous-readout SPAD arrays for ISM experiments [47]. Subsequently, we developed the 0.16 µm–BCD SPADs, with better PDE, narrower temporal response and lower afterpulsing probability, but with higher noise that can be lowered by cooling down the device to approximately 273 K, at the expense of an increased system complexity and a slightly higher afterpulsing probability.

Even if the final fill-factor is substantially higher than SPAD imagers with in-pixel electronics, it could be further improved by a micro-lenses array mounted on top of the detector. Custom-shaped micro-lenses arrays (MLA) can be deposited directly on the chip by exploiting recently-developed additive manufacturing techniques [50], with a theoretical equivalent fill-factor of more than 78% (i.e., higher than the value theoretically achievable with circular lenses). For applications requiring wider arrays, the pixel number can be increased up to several tens of elements, but the final limitation would be related to chip size (due to the high number of independent output pads) and to signal integrity constraints inside the chip itself (due to the external positioning of the readout/quenching circuits needed in order to maximize the fill-factor).

The replacement of the standard single-point detector with our SPAD array module can easily and reversibly transform any existing confocal laser-scanning microscope into an image scanning microscope, preserving all CLSM advantages (like the optical-sectioning capability) and without any need for prior calibrations. Furthermore, the single-photon timing capability of SPADs allows to add a further dimension to the measurements, making possible to combine ISM with FLIM, thus enabling straightforward FLISM experiments [47].

The ability of our SPAD array to image the excitation region (i.e. the detection volume) of a laser-scanning microscopy system can also improve the information content of many other advanced fluorescence microscopy techniques, such as fluorescence correlation spectroscopy. As an example, it can simultaneously measure the diffusion coefficients of a biomolecule for different detection volumes, giving access to the so-called FCS diffusion law [51]. In particular, by summing the fluorescence signal collected from different combination of pixels of the detector, it is possible to obtain virtual pinholes, thus detection volumes with different size [52]. Finally, from the analysis of the FCS diffusion law, it is possible to distinguish different biomolecular diffusion modes. In the context of FCS measurements, the single-photon timing capability of SPAD arrays will also open the way to a straightforward and synergic combination with fluorescence lifetime (FLCS), further improving the information content.

## 6. Conclusion

We designed, characterized ad evaluated two SPAD-based image sensors, specifically conceived for image-scanning microscopy applications. They are composed by 25 square pixels, having side dimension of 50 µm with a maximum achievable fill-factor higher than 50% (with further improvements made possible using micro-lenses). Each pixel integrates dedicated electronics for SPAD operation and is able to operate asynchronously from each other. Photon detections are marked with digital voltage pulses onto 25 independent output lines, with a time uncertainty lower than 200 ps (FWHM) at rates higher than $4\cdot 10^7$ events/s. Depending on the fabrication technology, performance can be directed towards high detection efficiency (more than 60% at 500 nm and about 30% at 650 nm) or low dark-count noise (200 cps at 25 °C) with low optical crosstalk probability (below 2%). Both image sensors are hosted into a standalone detection system, which has been used to validate them in ISM experiments, showing superior improvements upon simple confocal microscopy and also enabling fluorescence-lifetime, two-photon excitation, and stimulated emission depletion ISM implementations [47, 48, 27].

**Funding**

This work was partially supported by the European Union's Horizon 2020 research and innovation program under G.A. 731877 (SOLUS, an initiative of the Photonics Public Private Partnership) and G.A. 818699 (BrightEyes, a ERC Consolidator grant).